\documentclass[aps,superscriptaddress,preprintnumbers,showpacs]{revtex4}
\usepackage{amssymb}
\usepackage{dcolumn}
\usepackage{float}
\usepackage{bm}
\usepackage{amsmath}
\usepackage{mathrsfs}
\usepackage{slashed}
\usepackage{graphicx}
\begin{document}

\newcommand*{\PKU}{School of Physics and State Key Laboratory of Nuclear Physics and Technology, Peking University, Beijing 100871, China}\affiliation{\PKU}
\newcommand*{\chep}{Center for High Energy Physics, Peking University, Beijing 100871, China}\affiliation{\chep}

\title{Wave Packet Approach to Neutrino Oscillations with Matter Effects}

\author{Nan Qin}\affiliation{\PKU}
\author{Bo-Qiang Ma}\email{mabq@pku.edu.cn}\affiliation{\PKU}\affiliation{\chep}

\begin{abstract}
The complete description of neutrino oscillations require the wave
packet treatment. For terrestrial experiments the contribution from
the interaction of neutrino with the Earth matter would modify
oscillation pattern, i.e., the dependence of the flavor transition
probability on baseline $L$ and energy $E$. We thus combine the wave
packet approach and the matter effects, in order to have more
accurate descriptions of neutrino oscillation. The general
expression for the transition probability of oscillations with
arbitrary numbers of neutrinos are derived. As an example the
two-neutrino oscillation is studied in detail.
\end{abstract}

\maketitle

\section{introduction}
Over the past few decades the existence of neutrino oscillations has
been confirmed in a number of experiments. It is widely accepted
that neutrinos are massive and mixing. 
Neutrino oscillations are governed by the mass square differences
$\Delta m_{ji}^2$ of neutrino mass eigenstates and the neutrino
mixing matrix $U_{\alpha j}$ proposed by Pontecorvo, Maki, Nakawaga
and Sakata (PMNS)~\cite{Pontecorvo1957,Maki1962}. The standard
expression for the probability of the flavor transition
$\nu_{\alpha}\rightarrow\nu_{\beta}$ in vacuum is
\begin{align}
P_{\alpha\beta}(\vec{L})&=\sum_j|U_{\alpha j}|^2|U_{\beta j}|^2+2\textrm{Re}\sum_{j>i}U_{\alpha i}U_{\alpha j}^{\ast}U_{\beta i}^{\ast}U_{\beta j} \exp[-i(E_j-E_i)L]\nonumber\\
&\simeq\sum_j|U_{\alpha j}|^2|U_{\beta j}|^2+2\textrm{Re}\sum_{j>i}U_{\alpha i}U_{\alpha j}^{\ast}U_{\beta i}^{\ast}U_{\beta j}\exp[-i\frac{\Delta m_{ji}^2 L}{2E}],
\label{standard}
\end{align}
where $E$ and $\vec{L}$ are, respectively, the average energy of the
neutrino beam and the location of neutrino detector with respect to
the neutrino source. The $L/E$ dependence here is the most important
signature of neutrino oscillations which have been observed in
solar, atmospheric, accelerator and reactor neutrino oscillation
experiments. However, Eq.~(\ref{standard}) is a plane-wave
approximation and the complete derivation would require the use of
the wave packet formalism for the evolution of the massive neutrino
states. As discussed in
refs.~\cite{Kayser1981,Giunti2002,Akhmedov2010}, neutrino
oscillations are observable only if the process of neutrino
production and detection have momentum uncertainties that satisfy
the condition of coherent production and detection of different
neutrino mass eigenstates. Otherwise, different neutrino mass
components get decoherent and neutrino oscillations are destroyed.
Therefore from the quantum mechanical uncertainty principle,
neutrino states are naturally described by wave packets rather than
the plane wave with definite momentum. It is revealed that the
effects due to such wave packet treatment are observable in
oscillations with the existence of sterile
neutrinos~\cite{Hernandez2012} or oscillation with sufficiently long
baseline~\cite{Kayser2010}. In such cases decoherence of different
neutrino mass eigenstates modify the energy and distance dependence
of the flavor transition probability, as one can find
in~\cite{Giunti1998,Giunti2002}
\begin{align}
P_{\alpha\beta}(\vec{L})=\sum_j|U_{\alpha j}|^2|U_{\beta j}|^2+2\textrm{Re}\sum_{j>i}U_{\alpha i}U_{\alpha j}^{\ast}U_{\beta i}^{\ast}U_{\beta j} \exp[-2\pi i\frac{L}{L_{ji}^{\textrm{osc}}}-(\frac{L}{L_{ji}^{\textrm{coh}}})^2-2\pi^2(\frac{\sigma_x}{L_{ji}^{\textrm{osc}}})^2],\label{vacuum}
\end{align}
where $L_{ji}^{\textrm{osc}}\equiv\frac{4\pi E}{\Delta m_{ji}^2}$ is
the oscillation length,
$L_{ji}^{\textrm{coh}}\equiv\frac{4\sqrt{2}E^2\sigma_x}{\Delta
m_{ji}^2}$ is the coherence length and
$\sigma_x^2\equiv\sigma_{xP}^2+\sigma_{xD}^2$ is the quadratic sum
of the production localization and the detection localization.
Compared with the standard expression Eq.~(\ref{standard}), the two
damped terms, $-(\frac{L}{L_{ji}^{\textrm{coh}}})^2$ and
$-2\pi^2(\frac{\sigma_x}{L_{ji}^{\textrm{osc}}})^2$ in the
exponential, can be understood, respectively, as the decoherence due
to the separation of different mass eigenstates during propagation,
and as the decoherence due to the discrimination between different
mass eigenstates when neutrinos are emitted and/or absorbed. As any
of these two terms increases, the oscillation gets suppressed and in
the limit, Eq.~(\ref{vacuum}) reduces to the averaged transition
probability $\bar{P}_{\alpha\beta}(\vec{L})=\sum_j|U_{\alpha
j}|^2|U_{\beta j}|^2$ with no more $L/E$ dependence in it.

In terrestrial oscillation experiments, neutrino beams travel through the mantle and/or the core of the earth. The interaction of neutrinos with the particles of matter can change the pattern of oscillations, since the Hamiltonian of the neutrinos in matter is different from vacuum. In the flavor basis $(\nu_e,\nu_{\mu},\cdots)^T$, the effective Hamiltonian is given by
\begin{align}
H=U\textrm{diag}(E_1,E_2,\cdots)U^{\dagger}+\textrm{diag}(V,0,0,\cdots),\label{hamiltonian}
\end{align}
with $E_i\equiv\sqrt{m_i^2+|\vec{p}|^2}$ is energy of neutrino mass state $|\nu_i(\vec{p})\rangle$ in vacuum and $V\equiv\sqrt{2}G_f n_e$ ($n_e$ is the electron density in matter) is the charged-current contribution to the matter-induced effective potential of $\nu_e$~\cite{Wolfenstein1978,Barger1980}. Note that the neutral-current interactions are disregarded here because they are uniform in the flavor basis thus do not affect oscillations. Denoting the mixing matrix and Hamiltonian eigenvalues in matter with $\tilde{U}$ and $\tilde{E}_j$ respectively, we have
\begin{align}
\tilde{U}^{\dagger}H\tilde{U}=\textrm{diag}(\tilde{E}_1,\tilde{E}_2,\cdots).
\end{align}
Straightforwardly, one could find that the transition probability in matter is simply given by
\begin{align}
\tilde{P}_{\alpha\beta}(\vec{L})=P_{\alpha\beta}(\vec{L},U\rightarrow\tilde{U},E_j\rightarrow\tilde{E}_j),
\end{align}
in the plane-wave approximation. A natural question arises here: what about the complete wave packet approach with the existence of matter effects? We will look into it in the following sections.

\section{oscillations with arbitrary number of neutrinos}
It is well known that neutrinos are produced and detected in flavor eigenstates. Considering a production process\footnote{In some cases there are more than one initial particles or the final particle $P_F$ is absent, such as two-body weak decay $\pi^+\rightarrow\mu^+ +\nu_{\mu}$. But the following discussions are not affected.}
\begin{align}
P_I\rightarrow P_F+l_{\alpha}^{+}+\nu_{\alpha},\label{production}
\end{align}
the final neutrino state is generally described with wave packets as
\begin{align}
|\nu_{\alpha}\rangle=\int d^3 \vec{p}f_P(\vec{p},\vec{P}_{P})|\nu_{\alpha}(\vec{p})\rangle,
\end{align}
in which $f_P(\vec{p},\vec{P}_{P})$ is the momentum distribution function and $|\nu_{\alpha}(\vec{p})\rangle$ is the neutrino state with definite momentum $\vec{p}$. Usually $f_P(\vec{p},\vec{P}_{P})$ takes the gaussian form
\begin{align}
f_P(\vec{p},\vec{P}_{P})\propto \exp[-\frac{(\vec{p}-\vec{P}_P)^2}{4\sigma_P^2}]\label{gaussproduction}
\end{align}
which are sharply peaked around the average momentum $\vec{P}_P$ with uncertainty $\sigma_P$. For process described by Eq.~(\ref{production}) we have~\cite{Giunti2002}
\begin{align}
\vec{P}_P=\vec{P}_{P_I}-\vec{P}_{P_F}-\vec{P}_{l^+},~~~~~\sigma_P^2=\sigma_{P_I}^2+\sigma_{P_F}^2+\sigma_{l^+}^2.\label{parameter}
\end{align}
The first equation is the consequence of momentum conservation and
the second one tells that the effective momentum uncertainty of the
production is dominated by the particle with the largest momentum
uncertainty.

Similarly to the production process presented above, at $(\vec{L},T)$ with respect to the source, neutrinos are detected as wave packets
\begin{align}
|\nu_{\beta}\rangle=\int d^3\vec{p}f_D(\vec{p},\vec{P}_D)|\nu_{\beta}(\vec{p})\rangle\label{detectedstate}
\end{align}
with momentum distribution
\begin{align}
f_D\propto\exp[-\frac{(\vec{p}-\vec{P}_D)^2}{4\sigma_D^2}],\label{gaussdetection}
\end{align}
in which $\vec{P}_D$ and $\sigma_D$ has analogous definition as in
Eq.~(\ref{parameter}). In order to get the flavor transition
amplitude
$A_{\alpha\beta}(\vec{L},T)\equiv\langle\nu_{\beta}|\nu_{\alpha}(\vec{L},T)\rangle$,
we need to perform the space-time evolution operator
$\hat{S}\equiv\exp(-i\hat{H}T+i\hat{\vec{p}}\cdot\vec{L})$ upon the
initial neutrino state $|\nu_{\alpha}\rangle$. Here the Hamiltonian
operator $\hat{H}$ takes the form of Eq.~(\ref{hamiltonian}) such
that matter induced potential $V$ is taken into account. In most of
terrestrial oscillation experiments, it was shown
that~\cite{Krastev1988,Petcov1998,Chizhov1998} the relatively little
changes of electron density $n_e$ along the trajectories of
neutrinos crossing the Earth mantle or the mantle and the core are
neglected when the oscillation probabilities are calculated, thus
the constant density approximation
$n_e^{\textrm{man}(\textrm{core})}\equiv
\bar{n}_e^{\textrm{man}(\textrm{core})}$ with
$\bar{n}_e^{\textrm{man}(\textrm{core})}$ defined as mean electron
density number in the mantle (core) is sufficiently accurate.
Therefore the evolution operator $\hat{S}$ is invariant as neutrinos
traveling through the Earth and the neutrinos arriving at the
detector are described by~\footnote{For neutrinos crossing both the
mantle and the core one needs to divide $\hat{S}$ into
$\hat{S}^{\textrm{man}}\hat{S}^{\textrm{core}}\hat{S}^{\textrm{man}}$
with $n_e$ being either $\bar{n}_e^{\textrm{man}}$ or
$\bar{n}_e^{\textrm{core}}$ and calculate along the trajectory. The
derivation and expression for $A_{\alpha\beta}(\vec{L},T)$ is
analogous to what follows.}
\begin{align}
|\nu_{\alpha}(\vec{L},T)\rangle&=\int d^3 \vec{p}f_P(\vec{p},\vec{P}_{P})\exp(-i\hat{H}T+i\hat{\vec{p}}\cdot\vec{L})|\nu_{\alpha}(\vec{p})\rangle\nonumber\\
&=\int d^3\vec{p}f_P(\vec{p},\vec{P}_P)\sum_j\tilde{U}_{\alpha j}^{\ast}\exp(-i\tilde{E}_jT+i\vec{p}\cdot\vec{L})|\tilde{\nu}_j(\vec{p})\rangle,
\label{arrivedstate}
\end{align}
in which $|\tilde{\nu}_j(\vec{p})\rangle$ denotes the eigenstates of
Hamiltonian~(\ref{hamiltonian}). Expanding the detected neutrino
state $|\nu_{\beta}\rangle$ in Eq.~(\ref{detectedstate}) with the
same eigenstates and making use of the normalization condition
$\langle\tilde{\nu}_i(\vec{k})|\tilde{\nu}_j(\vec{p})\rangle=\delta_{ij}\delta^{(3)}(\vec{p}-\vec{k})$
we get
\begin{align}
A_{\alpha\beta}(\vec{L},T)&=\langle\nu_{\beta}|\nu_{\alpha}(\vec{L},T)\rangle\nonumber\\
&=\int d^3\vec{p} \int d^3\vec{k}f_P(\vec{p},\vec{P}_P)f_D^{\ast}(\vec{k},\vec{P}_D)\sum_{i,j}\tilde{U}_{\alpha j}^{\ast}\tilde{U}_{\beta i}\exp(-i\tilde{E}_j T+i\vec{p}\cdot\vec{L})\langle\tilde{\nu}_i(\vec{k})|\tilde{\nu}_j(\vec{p})\rangle\nonumber\\
&=\int d^3\vec{p}f_P(\vec{p},\vec{P}_P)f^{\ast}_D(\vec{p},\vec{P}_D)\sum_j \tilde{U}_{\alpha j}^{\ast}\tilde{U}_{\beta j}\exp(i\tilde{E}_j T+i\vec{p}\cdot\vec{L})
\end{align}
For the gaussian distributions Eq.~(\ref{gaussproduction}) and
Eq.~(\ref{gaussdetection}) it is easy to find that the overall
momentum distribution still takes the gaussian form with mean
momentum
$\vec{P}=\frac{\sigma_D^2\vec{P}_P+\sigma_P^2\vec{P}_D}{\sigma_P^2+\sigma_D^2}$
and momentum uncertainty
$\frac{1}{\sigma^2}=\frac{1}{\sigma_P^2}+\frac{1}{\sigma_D^2}$.
Notice that both $\vec{P}$ and $\sigma$ is dominated by the process
with the smaller momentum uncertainty. This is due to the fact that
a set of successive physical processes requires an overlap of the
wave packets in momentum space of all the processes, thus the one
with smallest momentum uncertainty determines the location and shape
of the overall wave packet. Therefore the amplitude reduces to
\begin{align}
A_{\alpha\beta}(\vec{L},T)\propto\int d^3\vec{p}\exp[-\frac{(\vec{p}-\vec{P})^2}{4\sigma^2}]\sum_j \tilde{U}_{\alpha j}^{\ast}\tilde{U}_{\beta j}\exp(-i\tilde{E}_jT+i\vec{p}\cdot\vec{L}).\label{amplitude}
\end{align}
Unlike the vacuum mixing matrix elements $U_{\alpha j}$ which are
constants, $\tilde{U}_{\alpha j}$ are functions of the neutrino
momenta $\vec{p}$. However, since the distribution function in
Eq.~(\ref{amplitude}) is sharply peaked around $\vec{P}$, it is a
good approximation to let $\tilde{U}_{\alpha j}=\tilde{U}_{\alpha
j}|_{\vec{p}=\vec{P}}$ and take them out of the integration over
$\vec{p}$. Consequently the integration can be performed with a
saddle point approximation around $\vec{P}$ leading to
\begin{align}
A_{\alpha\beta}(\vec{L},T)\propto\sum_j \tilde{U}_{\alpha j}^{\ast}\tilde{U}_{\beta j}\exp[-i\tilde{E}_j^0 T+i\vec{P}\cdot\vec{L}-\frac{(\vec{L}-\tilde{\vec{v}}_jT)^2}{4\sigma_x^2}],
\end{align}
in which $\tilde{E}_j^0\equiv\tilde{E}_j|_{\vec{p}=\vec{P}}$ and $\tilde{\vec{v}}_j=\frac{\partial\tilde{E}_j}{\partial\vec{p}}|_{\vec{p}=\vec{P}}$ are, respectively, the mean energy and mean group velocity of the corresponding wave packet in matter. The space-time uncertainty $\sigma_x\equiv\frac{1}{2\sigma}$ describes localization of both the production and detection process because of the relation $\sigma_x^2=\sigma_{xP}^2+\sigma_{xD}^2$ with $\sigma_{xP}\equiv\frac{1}{2\sigma_P}$ and $\sigma_{xD}\equiv\frac{1}{2\sigma_D}$ are space-time uncertainties of the two processes respectively. Notice that opposite to momentum uncertainties, the overall space-time uncertainty is dominated by the larger one of the localizations.

In a majority of neutrino oscillation experiments propagation time $T$ is unmeasured. Even when $T$ is accurately measured, the most important observable is the events number accumulated during a period of time. Therefore rather than the transition probability $P_{\alpha\beta}(\vec{L},T)\equiv|A_{\alpha\beta}(\vec{L},T)|^2$, we are interested in the time average of it: $P_{\alpha\beta}(\vec{L})\propto\int dT|A_{\alpha\beta}(\vec{L},T)|^2$ with the normalization condition $\sum_{\alpha} P_{\alpha\beta}(\vec{L})=1$. After integration over $T$ we get
\begin{align}
P_{\alpha\beta}(\vec{L})&\propto\sum_{ij}\tilde{U}_{\alpha i}\tilde{U}_{\alpha j}^{\ast}\tilde{U}_{\beta i}^{\ast}\tilde{U}_{\beta j}\sqrt{\frac{4\pi\sigma_x^2}{\tilde{v}_i^2+\tilde{v}_j^2}}\exp[\frac{\sigma_x^2}{\tilde{v}_i^2+\tilde{v}_j^2} \left(-i(\tilde{E}_j^0-\tilde{E}_i^0)+\frac{\vec{L}\cdot(\tilde{\vec{v}}_j+\tilde{\vec{v}}_i)}{2\sigma_x^2}\right)^2]\nonumber\\
&\propto\sum_{ij}\tilde{U}_{\alpha i}\tilde{U}_{\alpha j}^{\ast}\tilde{U}_{\beta i}^{\ast}\tilde{U}_{\beta j}\sqrt{\frac{4\pi\sigma_x^2}{\tilde{v}_i^2+\tilde{v}_j^2}}\exp[-i(\tilde{E}_j^0-\tilde{E}_i^0)\vec{L}\cdot \frac{\tilde{\vec{v}}_j+\tilde{\vec{v}}_i}{\tilde{v}_i^2+\tilde{v}_j^2}-\frac{\sigma_x^2}{2}(\tilde{E}_j^0-\tilde{E}_i^0)^2 \frac{2}{\tilde{v}_i^2+\tilde{v}_j^2}+\frac{(\vec{L}\cdot\tilde{\vec{v}}_j+\vec{L}\cdot\tilde{\vec{v}}_i)^2}{4\sigma_x^2(\tilde{v}_i^2+ \tilde{v}_j^2)}].
\label{generalprobability}
\end{align}
Let us have a look at some features of Eq.~(\ref{generalprobability}), compared with the probability in vacuum, Eq.~(\ref{vacuum}). The first is, as expected, the replacement of mixing matrix elements $U_{\alpha j}\rightarrow\tilde{U}_{\alpha j}$. The square root term can be factored out of the summation in relativistic limit thus do not contribute after the normalization. Second, the eigenenergies are changed from $E_j^0$ to $\tilde{E}_j^0$ which is also naturally expected. Moreover, all terms in the exponential are multiplied by combinations of the group velocities of the wave packets corresponding to the matter eigenstates $|\tilde{\nu}_j\rangle$, as a result, the oscillation pattern might be modified depending on the magnitude of matter effect and neutrino energy. In the next section we will discuss the two-neutrino oscillation as a specific example of our formalism and then, it will be much clearer about how such corrections influence the dependence of transition probabilities on $L$ and $E$.

\section{two-neutrino oscillation}
In general the number of massive neutrinos can be larger than 3, for instance, the existence of sterile neutrinos is largely discussed (see~\cite{Giunti2011} for a brief review). However, all compelling data on neutrino oscillations can still be described with three light neutrinos. In a large proportion of experiments, two-flavor neutrino oscillations serve as good approximations to the three-neutrino description. Therefore next we discuss the flavor transition between $\nu_e$ and $\nu_{\mu}$ within the formalism presented in Sec.~II. The Hamiltonian~(\ref{hamiltonian}) for $\nu_e$-$\nu_{\mu}$ system is given by
\begin{align}
H=U\left(\begin{array}{ccc}
E_1&0\\
0&E_2
\end{array}\right)U^{\dagger}+\left(\begin{array}{ccc}V&0\\0&0\end{array}\right)
\end{align}
with
\begin{align}
U=\left(\begin{array}{ccc}
\cos{\theta}&\sin{\theta}\\
-\sin{\theta}&\cos{\theta}
\end{array}\right).
\end{align}
It is straightforward to diagonalize such a Hamiltonian and, consequently we get the eigenenergies and mixing matrix in matter as
\begin{align}
&\tilde{E}_1^0=\frac{1}{2}\left(-\sqrt{2 \left(E_1^0-E_2^0\right) V \cos {2 \theta
}-2 E_1^0 E_2^0+(E_1^0-E_2^0)^2+V^2}+E_1^0+E_2^0+V\right),\\
&\tilde{E}_2^0=\frac{1}{2}\left(\sqrt{2 \left(E_1^0-E_2^0\right) V \cos {2 \theta
}-2 E_1^0 E_2^0+(E_1^0-E_2^0)^2+V^2}+E_1^0+E_2^0+V\right),\\
&\tilde{U}=\left(
\begin{array}{cc}
 \frac{C}{\left(E_1^0-E_2^0\right) \sqrt{\frac{C^2}{\left(E_1^0-E_2^0\right){}^2}+1}} & -\frac{D}{\left(E_1^0-E_2^0\right) \sqrt{\frac{D^2}{\left(E_1^0-E_2^0\right){}^2}+1}} \\
 \frac{1}{\sqrt{\frac{C^2}{\left(E_1^0-E_2^0\right){}^2}+1}} & \frac{1}{\sqrt{\frac{D^2}{\left(E_1^0-E_2^0\right){}^2}+1}}
\end{array}
\right),
\end{align}
with definitions
\begin{align}
&E_j^0\equiv\sqrt{m_j^2+|\vec{P}|^2},\\
&C\equiv\left(-V+\left(E_2^0-E_1^0\right) \cos {2 \theta }+\sqrt{\left(E_1^0-E_2^0\right){}^2+2 V \cos {2 \theta } \left(E_1^0-E_2^0\right)+V^2}\right)
\csc{2 \theta},\\
&D\equiv\left(V+\left(E_2^0-E_1^0\right) \cos {2 \theta }+\sqrt{\left(E_1^0-E_2^0\right){}^2+2 V \cos {2 \theta } \left(E_1^0-E_2^0\right)+V^2}\right)
\csc{2 \theta}.
\end{align}
One can verify that our result above is equivalent to the redefinitions of mixing angle in matter $\theta_m$ and oscillation length in matter $l_M$ derived in Ref.~\cite{Wolfenstein1978,Barger1980}. According to the PREM model~\cite{Dziewonski1981}, the mean electron number densities in the mantle and the core in the Earth are respectively $\bar{n}_e^{\textrm{man}}\cong2.2~\textrm{cm}^{-3}\textrm{N}_\textrm{A}$ and $\bar{n}_e^{\textrm{core}}\cong5.4~\textrm{cm}^{-3}\textrm{N}_\textrm{A}$. So the magnitude of the effective potential inside the Earth is $V\sim10^{-13}\textrm{~eV}$, which is very small compared with the neutrino energy. Therefore we neglect terms of $\mathcal{O}(V^3)$. As a result, the appearance probability for the flavor transition $\nu_e\rightarrow\nu_{\mu}$ is obtained by substituting the eigenenergies $\tilde{E}_1^0$, $\tilde{E}_2^0$, and the mixing matrix elements $\tilde{U}_{\alpha j}$ into Eq.~(\ref{generalprobability}), which in the relativistic approximation leads to
\begin{align}
P_{e\mu}(\vec{L})=\sum_j|\tilde{U}_{ej}|^2|\tilde{U}_{\mu j}|^2+2\textrm{Re}\{\tilde{U}_{e 1}\tilde{U}_{e 2}^{\ast}\tilde{U}_{\mu 1}^{\ast}\tilde{U}_{\mu 2}\exp[-2\pi i\frac{L}{\tilde{L}_{21}^{\textrm{osc}}}-(\frac{L}{L_{21}^{\textrm{coh}}})^2-2\pi^2(\frac{\sigma_x}{\tilde{L}_{21}^{\textrm{osc}}})^2+(\frac{L}{L_{21}^{\textrm{mcoh}}})^2]\}. \label{pemusimplified}
\end{align}
Here we defined the oscillation length in matter
\begin{align}
\tilde{L}_{21}^{\textrm{osc}}\equiv\frac{2\pi}{\tilde{E}_2^0-\tilde{E}_1^0},
\end{align}
which is naturally expected and the matter-coherence length
\begin{align}
L_{21}^{\textrm{mcoh}}\equiv\frac{2\sqrt{2}E\sigma_x}{V\sin{2\theta}}.
\end{align}
The matter-coherence term in Eq.~(\ref{pemusimplified}), $\exp{(\frac{L}{L_{21}^{\textrm{mcoh}}})^2}$, will enhance the oscillation significantly when $L$ approaches $L_{21}^{\textrm{mcoh}}$, while the propagation-decoherence term $\exp[-(\frac{L}{L_{21}^{\textrm{coh}}})^2]$ suppresses the oscillation when $L$ approaches $L_{21}^{\textrm{coh}}$. Considering the magnitude of the mass square difference $\Delta m_{21}^2\sim10^{-5}\textrm{~eV}^2$, we find that $\frac{L_{21}^{\textrm{mcoh}}}{L_{21}^{\textrm{coh}}}\sim\frac{10^8\textrm{~eV}}{E}$, which tells that the matter-coherence term dominates over the propagation-decoherence term when the neutrino mean energy $E$ is much larger than $100 \textrm{~MeV} $ and vice versa. On the other hand we have $\frac{L_{21}^{\textrm{mcoh}}}{L_{21}^{\textrm{osc}}}\sim10^8 \textrm{~eV}\cdot\sigma_x$ and $\frac{L_{21}^{\textrm{coh}}}{L_{21}^{\textrm{osc}}}\sim E\sigma_x$, which means if we are interested in the matter-coherence effect and/or the propagation-decoherence effect near the first oscillation maximum, the localization of either production or detection process must be sufficiently small. And for small $\sigma_x$, effects due to the localization-decoherence term $\exp[-2\pi^2(\frac{\sigma_x}{\tilde{L}_{21}^{\textrm{osc}}})^2]$ can be totally neglected.

For oscillations of antineutrinos in matter, the transition probability can formally be obtained by replacing $V$ with $-V$ (and $\delta_{\textrm{CP}}$ with $-\delta_{\textrm{CP}}$ in general cases) in the corresponding equations, for instance $P_{\bar{e}\bar{\mu}}\equiv P_{e\mu}(V\rightarrow-V)$. Notice that the matter-coherence term in Eq.~(\ref{pemusimplified}) enhances both $P_{e\mu}$ and $P_{\bar{e}\bar{\mu}}$, different from the Mikheyev, Smirnov, Wolfenstein (or MSW) effect which leads to resonance enhancement either of $P_{e\mu}$ or $P_{\bar{e}\bar{\mu}}$ but not of both~\cite{Mikheev1985,Mikheev1986,Barger1980}. And due to the dependence of $\tilde{U}_{\alpha j}$ and $\tilde{E}_j$ on $V$ we have neither $P_{e\mu}= P_{\bar{e}\bar{\mu}}$ nor $P_{ee}= P_{\bar{e}\bar{e}}$. This is the result of the fact that matter in the Earth is not charge symmetric, thus the effective potential $V$ and consequently Hamiltonian~(\ref{hamiltonian}) is neither CP- nor CPT- invariant.

\begin{figure}[H]
\begin{center}
\includegraphics[width=9cm]{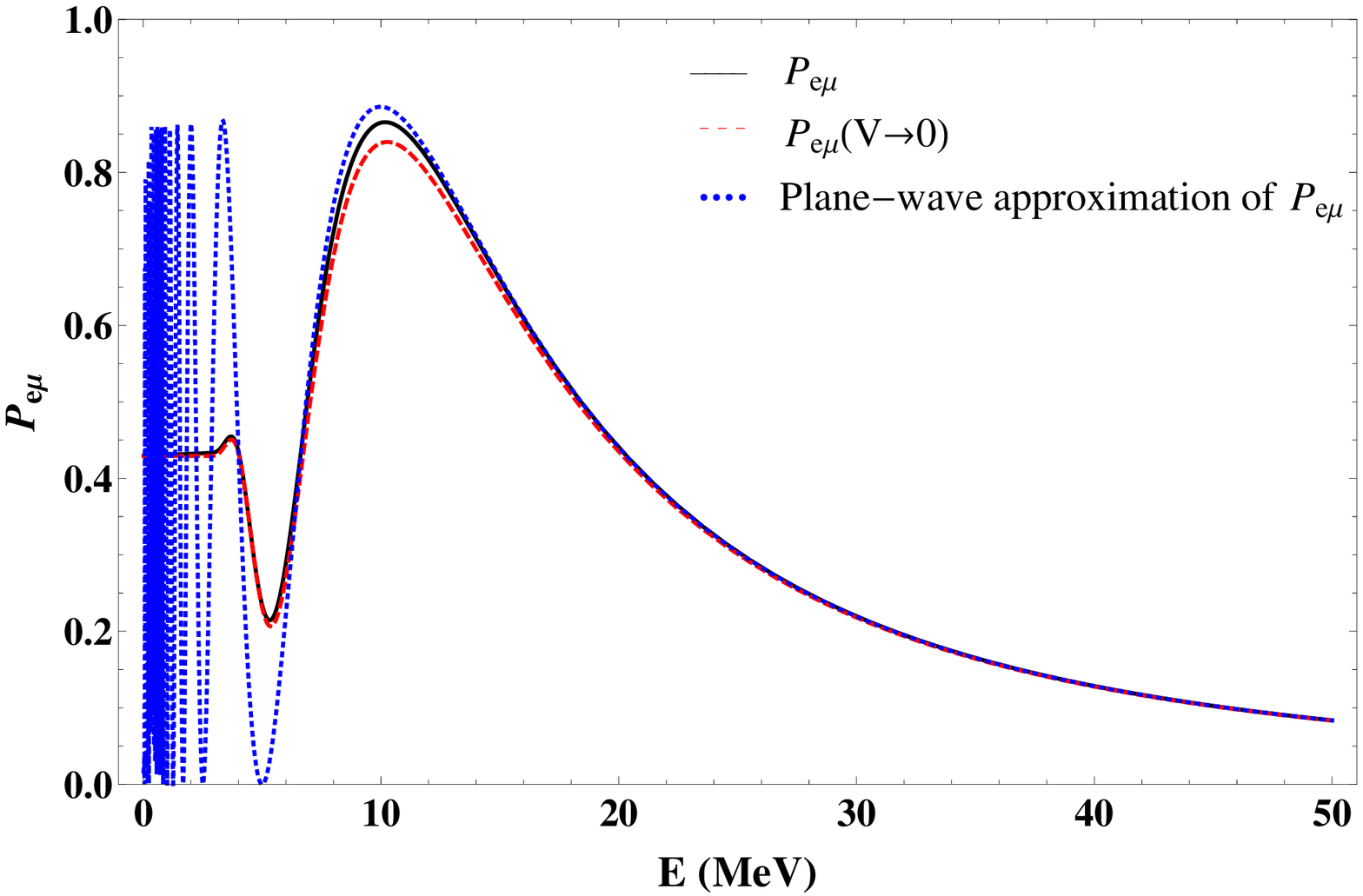}~~~~
\includegraphics[width=9cm]{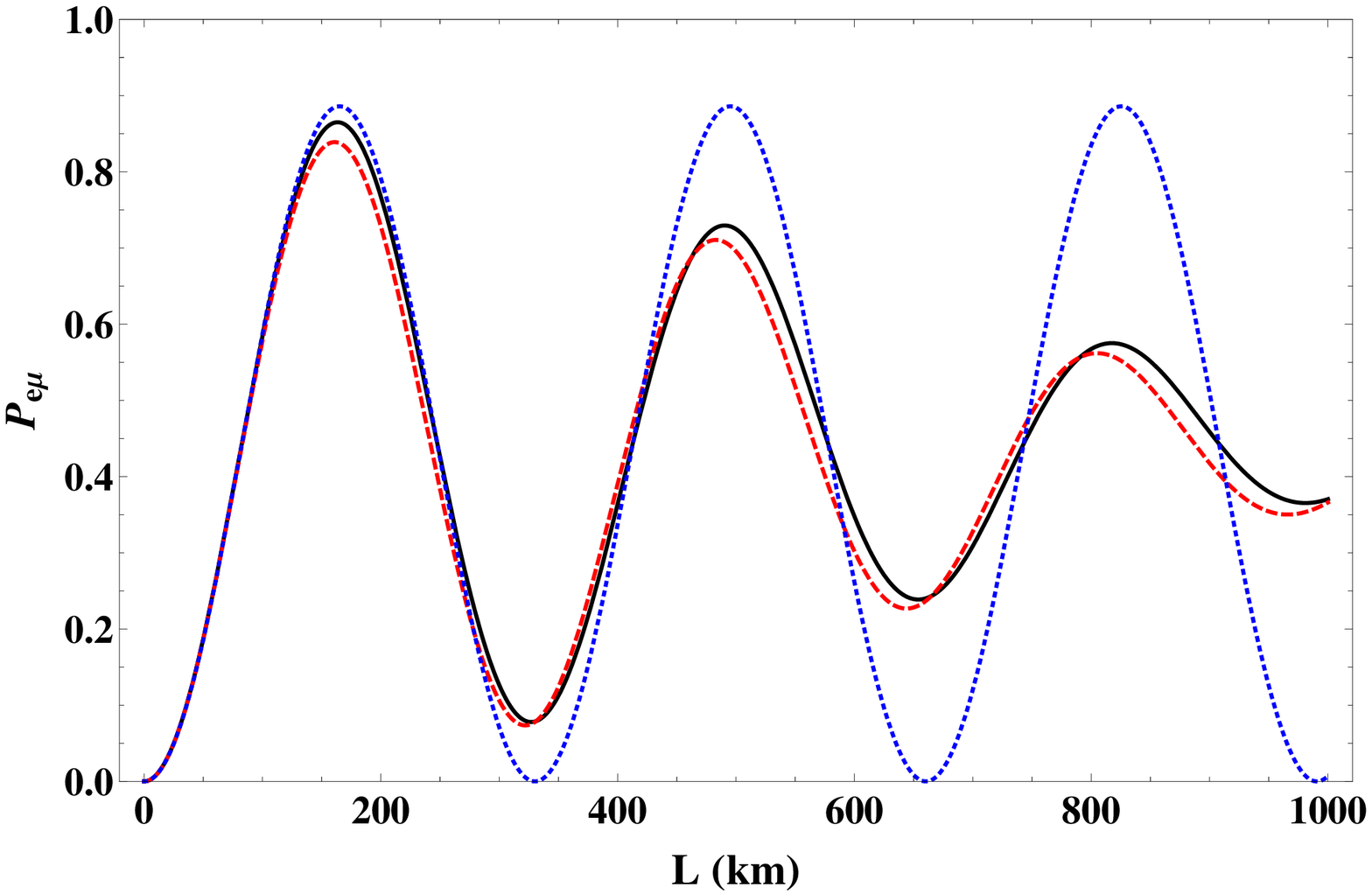}
\caption{Transition probability $P(\nu_e\rightarrow\nu_{\mu})$ with $\sigma_x=10^{-11}\textrm{~cm}$, $V=1.65\times10^{-13}\textrm{~eV}$, $\sin^2{\theta_{12}}=0.312$, $\Delta m_{21}^2=7.58\times10^{-5}\textrm{~eV}^2$ and $L=164 \textrm{~km}$ in the left panel while $E=10 \textrm{~MeV}$ in the right panel. The black solid line denotes the result from Eq.~(\ref{pemusimplified}), the blue dotted line denotes the result of the plane-wave approximation and the red dashed line denotes the result omitting the matter effect.}\label{figpemu}
\end{center}
\end{figure}

With the global best fit values of $\theta_{12}$, $\Delta m_{21}^2$ in Ref.~\cite{Fogli2011} and $V(n_e=\bar{n}_e^{\textrm{man}})$, the behavior of $P_{e\mu}(\vec{L})$ against $L$ and $E$ are illustrated in Fig.~(\ref{figpemu}).
Let us have a brief analysis of the curves presented above. First, the oscillation of plane-wave approximation is apparently larger than the other two. Besides the MSW effect, this is also because the localization $\sigma_x$ is set to be relatively small, so the decoherence effect due to $\exp[-(\frac{L}{L_{21}^{\textrm{coh}}})^2]$ in Eq.~(\ref{pemusimplified}) becomes important, and grows as $L$ increases and/or $E$ decreases. There are also horizontal shift of $P_{e\mu}(V\rightarrow0)$ from the other two, which is because of the difference between $L_{21}^{\textrm{osc}}$ and $\tilde{L}_{21}^{\textrm{osc}}$. Moreover, although the oscillation are both suppressed in $P_{e\mu}$ and $P_{e\mu}(V\rightarrow0)$, the matter-coherence effect due to $\exp[-(\frac{L}{L_{21}^{\textrm{mcoh}}})^2]$ in Eq.~(\ref{pemusimplified}) enhances the probability, such that $P_{e\mu}$ is slightly above $P_{e\mu}(V\rightarrow0)$ in Fig.~(\ref{figpemu}).

\begin{figure}[H]
\begin{center}
\includegraphics[width=9cm]{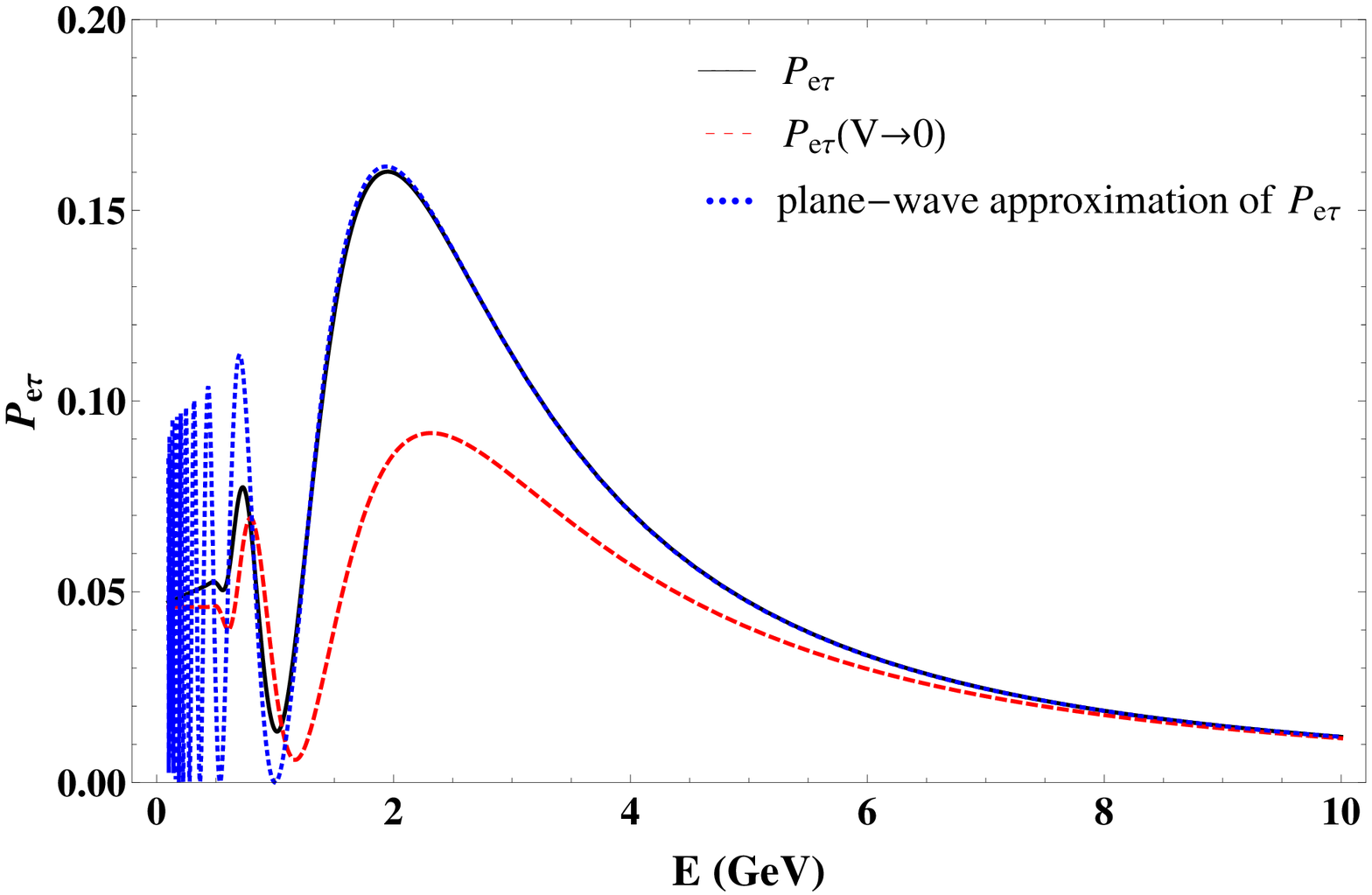}~~~~
\includegraphics[width=9cm]{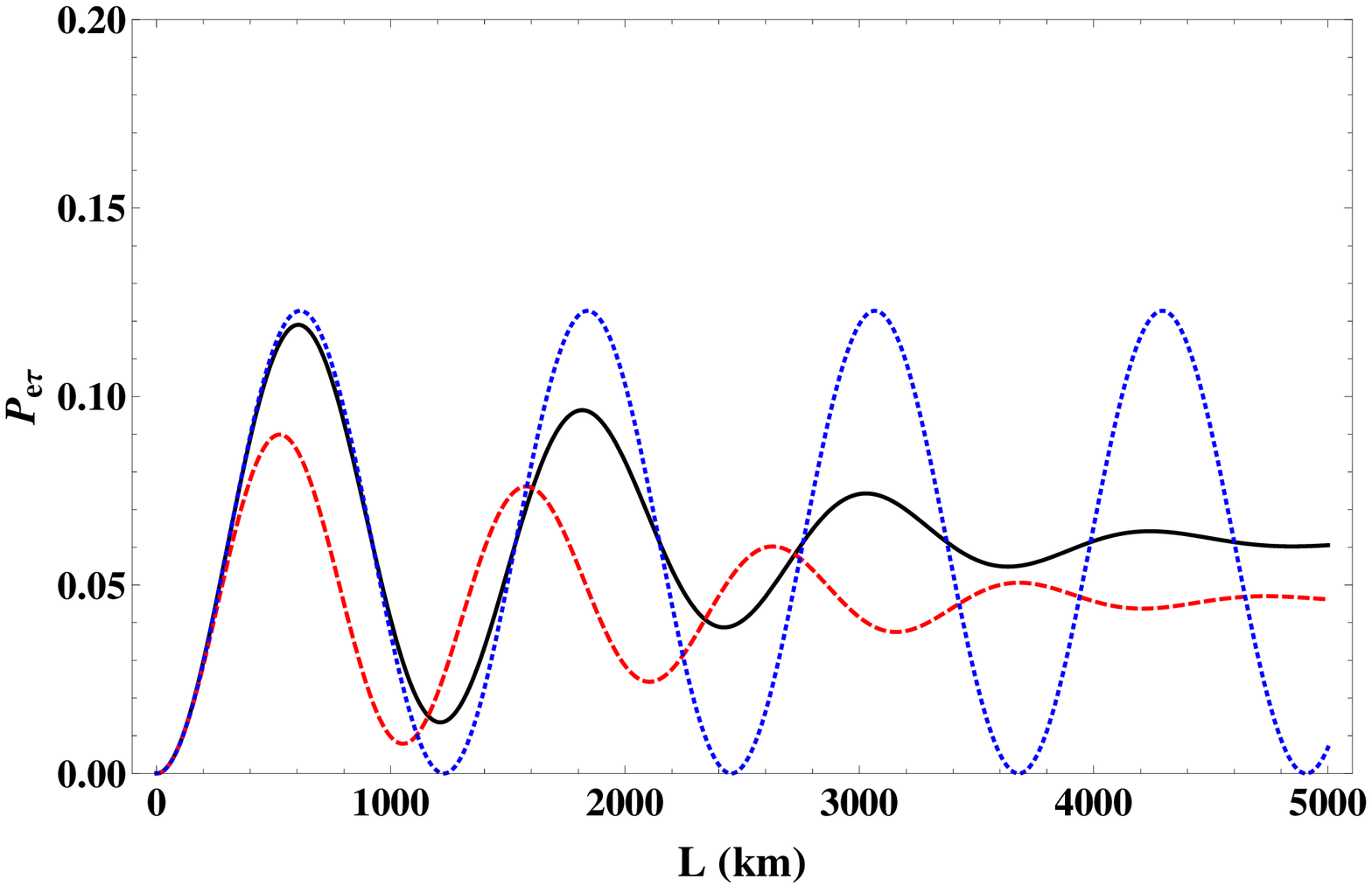}
\caption{Transition probability $P(\nu_e\rightarrow\nu_{\tau})$ with $\sigma_x=10^{-13}\textrm{~cm}$, $V=1.65\times10^{-13}\textrm{~eV}$, $\sin^2{2\theta_{13}}=0.092$, $\Delta m_{31}^2=2.32\times10^{-3}\textrm{~eV}^2$ and $L=1227 \textrm{~km}$ in the left panel while $E=1 \textrm{~GeV}$ in the right panel. The black solid line denotes the result from Eq.~(\ref{pemusimplified}), the blue dotted line denotes the result of the plane-wave approximation and the red dashed line denotes the result omitting the matter effect.}\label{figpetaunormal}
\end{center}
\end{figure}

\begin{figure}[H]
\begin{center}
\includegraphics[width=9cm]{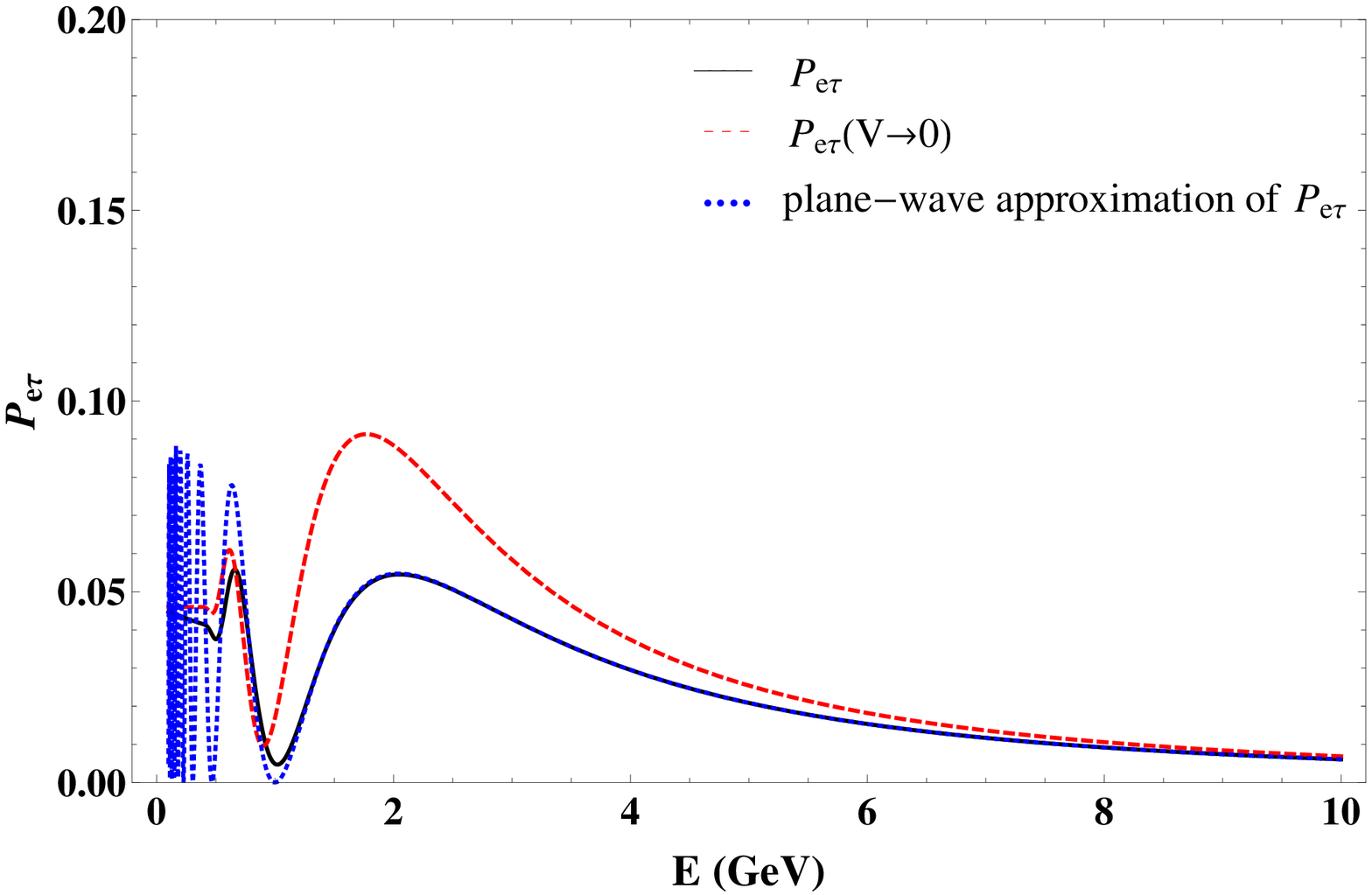}~~~~
\includegraphics[width=9cm]{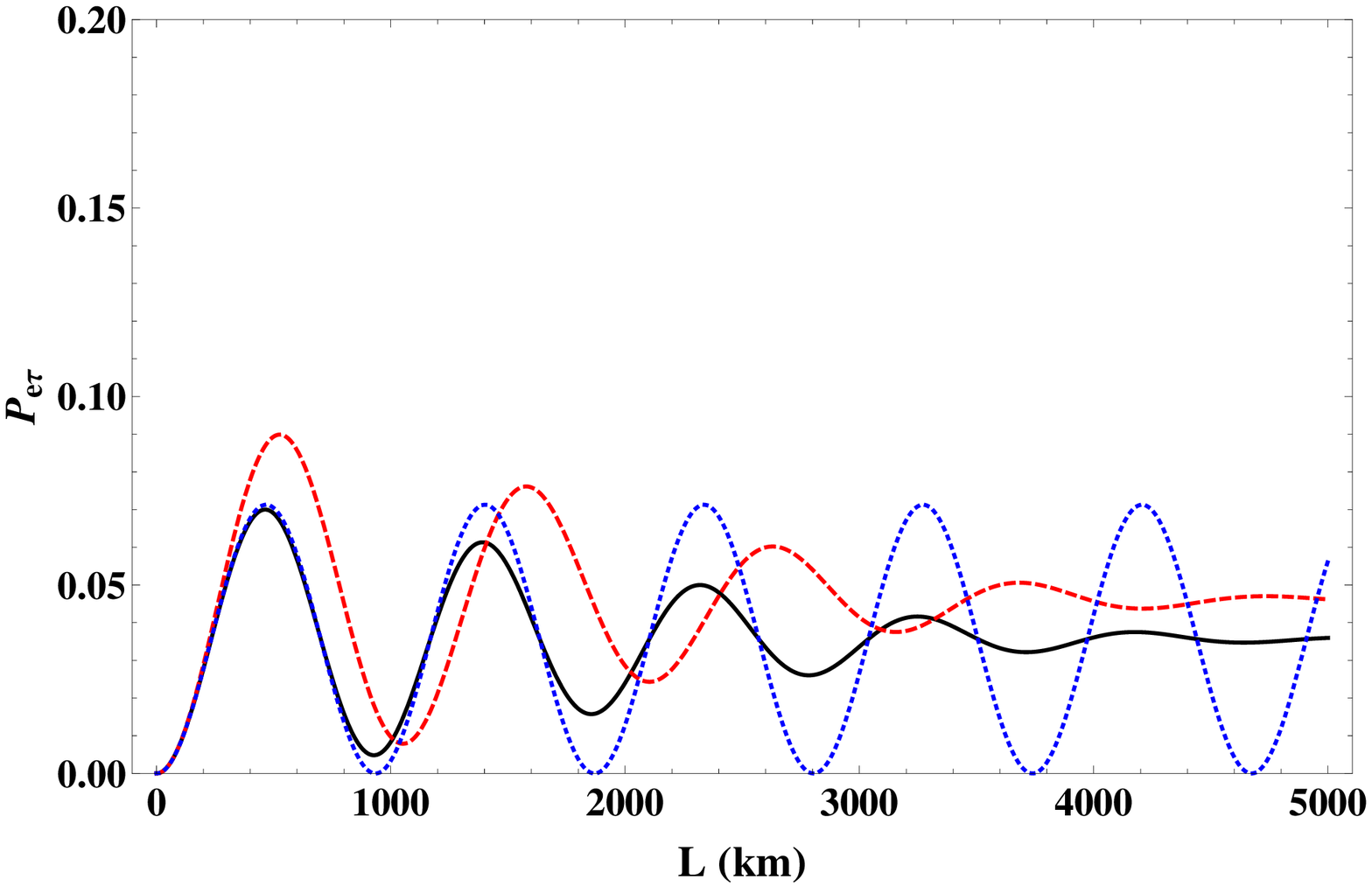}
\caption{Transition probability $P(\nu_e\rightarrow\nu_{\tau})$ with $\sigma_x=10^{-13}\textrm{~cm}$, $V=1.65\times10^{-13}\textrm{~eV}$, $\sin^2{2\theta_{13}}=0.092$, $\Delta m_{31}^2=-2.32\times10^{-3}\textrm{~eV}^2$ and $L=935 \textrm{~km}$ in the left panel while $E=1 \textrm{~GeV}$ in the right panel. The black solid line denotes the result from Eq.~(\ref{pemusimplified}), the blue dotted line denotes the result of the plane-wave approximation and the red dashed line denotes the result omitting the matter effect.}\label{figpetauinverted}
\end{center}
\end{figure}

For oscillation between $\nu_e$ and $\nu_{\tau}$ it is just to replace the mixing angle $\theta_{12}$ and mass-square difference $\Delta m_{21}^2$ in each terms of Eq.~(\ref{pemusimplified}) with $\theta_{13}$ and $\Delta m_{31}^2$. With the best-fit value of $\theta_{13}$ from the recent published Daya Bay experiment~\cite{An2012} and $|\Delta m_{31}^2|$ from the MINOS experiment~\cite{Adamson2011}, the appearance probability $P_{e\tau}$ of $\nu_e\leftrightarrow\nu_{\tau}$ oscillation with normal mass hierarchy $\Delta m_{31}^2>0$ is illustrated in Fig.~(\ref{figpetaunormal}) and the one with inverted hierarchy $\Delta m_{31}^2<0$ in Fig.~(\ref{figpetauinverted}).

Apparently, the MSW effect leads to the large enhancement in the case of normal mass hierarchy while large suppression in the case of inverted mass hierarchy. Besides, the mass-coherence effect and the difference between $L_{31}^{\textrm{osc}}$ and $\tilde{L}_{31}^{\textrm{osc}}$ contribute to both vertical and horizontal shift of $P_{e\tau}$ from $P_{e\tau}(V\rightarrow0)$. Similarly to the $P_{e\mu}$ case, as $L$ increases and/or $E$ decreases, contribution of propagation decoherence suppresses the oscillation exponentially. Notice that here for $\nu_e\leftrightarrow\nu_{\tau}$, we need a smaller localization, $\sigma_x\sim10^{-13}\textrm{~cm}$, to have apparent decoherence and matter-coherence effect. This is because $|\Delta m_{31}^2|$ and $E$ are both larger than the previous case, and in order to avoid $L_{31}^{\textrm{coh}}\gg L_{31}^{\textrm{osc}}$ and/or $L_{31}^{\textrm{mcoh}}\gg L_{31}^{\textrm{osc}}$, smaller localizations of both production and detection are necessary.

We conclude this section with some comments on the main results
Eq.~(\ref{pemusimplified}), Fig.~(\ref{figpemu}),
Fig.~(\ref{figpetaunormal}) and Fig.~(\ref{figpetauinverted}). The
two neutrino flavor transition probability with matter effects in
the wave packet formalism is accurately described by
Eq.~(\ref{pemusimplified}), as long as the matter induced effective
potential, $V$, is small, since we have ignored contributions of
$\mathcal{O}(V^3)$. Otherwise, the matter-coherence term
$\exp[(\frac{L}{L_{21}^{\textrm{mcoh}}})^2]$ would destroy the
unitarity of probability unless we include higher order terms of
$V$. In all the three figures, we present the behaviors of
$P_{e\mu(\tau)}$, $P_{e\mu(\tau)}(V\rightarrow0)$ and
$P_{e\mu(\tau)}$ with plane-wave approximation, from which we find
that if oscillation parameters ($L$, $E$ and especially $\sigma_x$)
are set properly, the predictions of wave packet approach with
matter effects are apparently different from the predictions either
without matter effects or without wave packet effects. Therefore, if
observations on the wave packet formalism are performed,
contributions of matter effects can not be trivially ignored. We
note here that the two-neutrino oscillations serve only as
approximations of the physical truth. For accurate data analysis, it
is required to work with three-neutrino oscillations which,
mathematically, is a little bit more complicated.

\section{discussions and conclusions}
In the wave packet formalism propagation decoherence is fundamentally due to the difference between the group velocities of the different mass components. For the two-neutrino oscillations with matter effects discussed in the previous section, we have
\begin{align}
v_1\simeq1-\frac{m_1^2}{2E^2}-\frac{V^2\sin^2{2\theta}}{2\Delta m_{21}^2},~~~~
v_2\simeq1-\frac{m_2^2}{2E^2}+\frac{V^2\sin^2{2\theta}}{2\Delta m_{21}^2}
\end{align}
for velocities with $\Delta m_{21}^2>0$. It is indicated that matter effects decrease the group velocity of the lighter neutrino wave packet while increase the group velocity of the heavier neutrino wave packet. Notice that when neutrino energy is sufficiently large, more terms need to be included and the velocities will be always less than $1$.

\begin{figure}[H]
\begin{center}
\includegraphics[width=9cm]{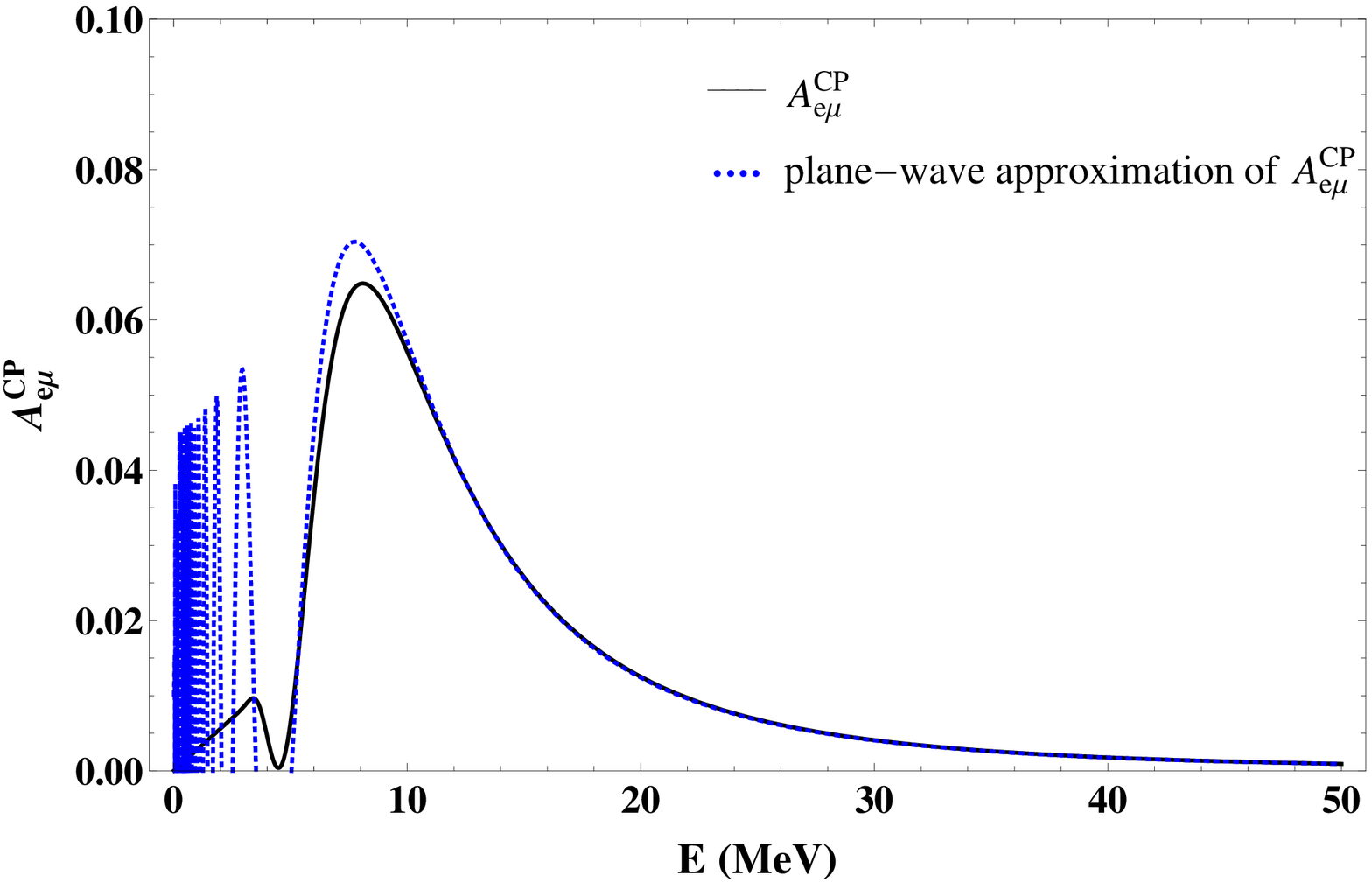}~~~~
\includegraphics[width=9cm]{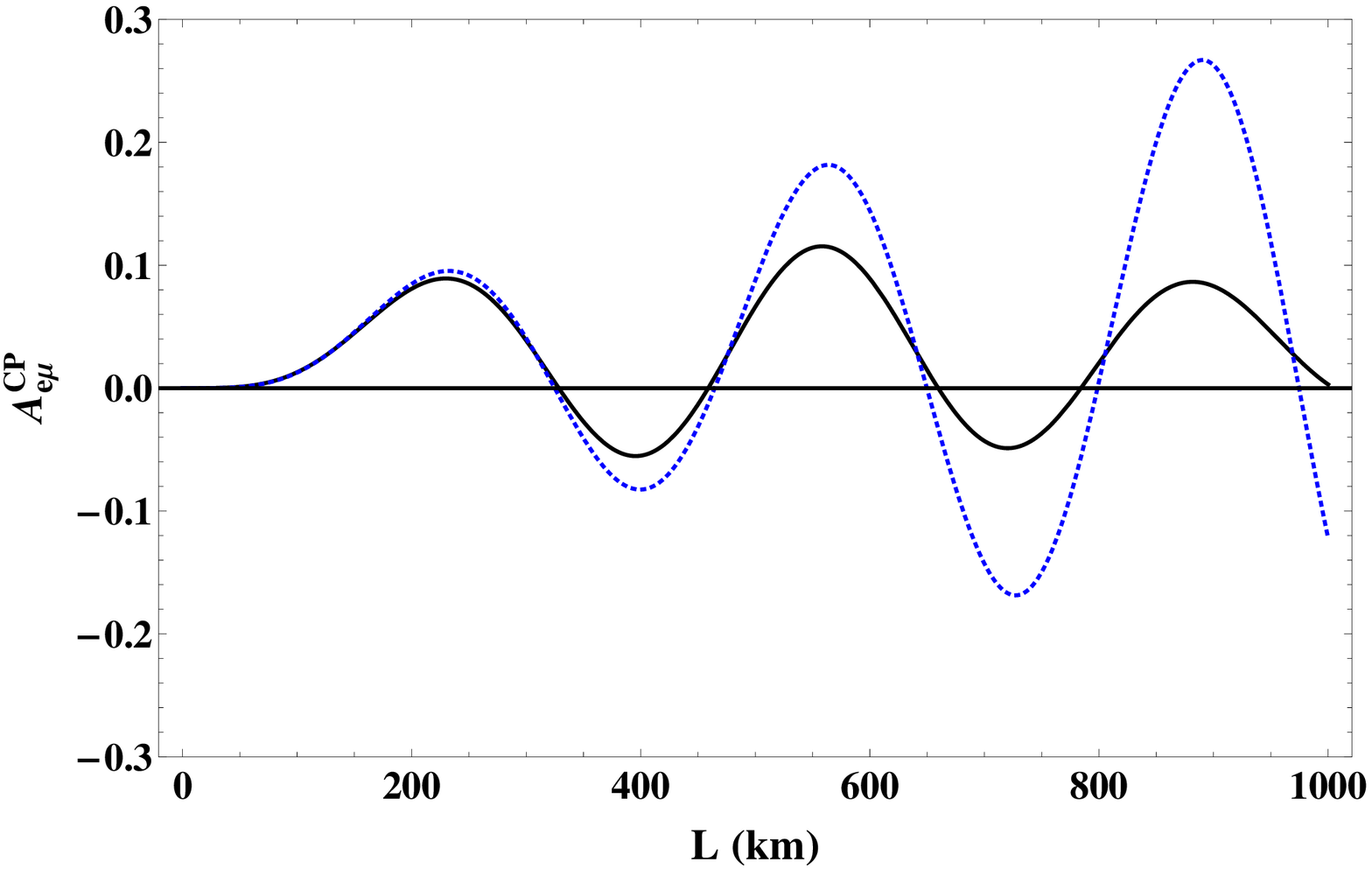}
\caption{CP asymmetry $A_{e\mu}^{\textrm{CP}}$ with $\sigma_x=10^{-11}\textrm{~cm}$, $V=1.65\times10^{-13}\textrm{~eV}$, $\sin^2{\theta_{12}}=0.312$, $\Delta m_{21}^2=7.58\times10^{-5}\textrm{~eV}^2$ and $L=164 \textrm{~km}$ in the left panel while $E=10 \textrm{~MeV}$ in the right panel. The black solid line denotes the result from Eq.~(\ref{pemusimplified}) and the blue dotted line denotes the result of the plane-wave approximation.}\label{asymmetry}
\end{center}
\end{figure}

Recent reactor oscillation experiments~\cite{An2012,Ahn2012} have revealed that $\theta_{13}$ is much larger than $0$, which makes the measurement of CP violation in neutrino oscillations possible and easier. Besides the contribution of the CP phase $\delta_{\textrm{CP}}$ in the mixing matrix, there are also CP violation due to the matter effects, even in the two-neutrino oscillation system as discussed in the previous section. As an example, defining CP asymmetry $A_{e\mu}^{\textrm{CP}}\equiv P_{e\mu}-P_{\bar{e}\bar{\mu}}$, Fig.~(\ref{asymmetry}) shows the behavior of $A_{e\mu}^{\textrm{CP}}$ compared with the plane-wave approximation.
As indicated in Fig.~(\ref{asymmetry}), when $L$ increases or $E$ decreases the difference between the wave packet result and the plane-wave approximation becomes more significant. Therefore, in order to measure CP violation precisely, it is necessary to work within the wave packet formalism in certain cases.

In conclusion, for terrestrial experiments, the wave packet description of neutrino oscillations, Eq.~(\ref{vacuum}), is modified by the interactions between neutrinos and the Earth matter they travel through, as indicated in Eq.~(\ref{generalprobability}) and Eq.~(\ref{pemusimplified}), which are the main results of this paper. From Fig.~(\ref{figpemu}), Fig.~(\ref{figpetaunormal}) and Fig.~(\ref{figpetauinverted}), we find that the complete wave packet approach with matter effects to neutrino oscillations gives distinctive predictions under certain parameter settings. Therefore, for future terrestrial neutrino oscillation experiments, especially with small localizations of neutrino emission and absorption, wave packet approach with matter effects can describe the data with more accuracy.

\begin{acknowledgements}
This work was supported by National Natural Science Foundation of China (Nos.~10975003, 11021092, 11035003 and 11120101004).
\end{acknowledgements}

\end{document}